\documentclass[letter]{aa} 

\usepackage{graphicx}
\usepackage{txfonts}
\usepackage{natbib}
\bibpunct{(}{)}{;}{a}{}{,}

\usepackage{lscape}
\usepackage{longtable}
\usepackage{amssymb}

\def\kms{km~s$^{-1}$}
\def\G23{G023.01$-$00.41}

\usepackage{color}

\begin{document}

  \title{Properties of a radio jet within 1000\,AU of an O-type YSO}
%
   \title{Momentum-driven outflow emission from an O-type YSO}
   \subtitle{Comparing the radio jet with the molecular outflow}
%

  \author{A. Sanna \inst{1} \and L. Moscadelli \inst{2} \and R. Cesaroni \inst{2}  \and A. Caratti~o~Garatti \inst{3}
  \and C. Goddi \inst{4}  \and C. Carrasco-Gonz\'alez \inst{5}}
             

   \institute{Max-Planck-Institut f\"{u}r Radioastronomie, Auf dem H\"{u}gel 69, 53121 Bonn, Germany\\
   \email{asanna@mpifr-bonn.mpg.de}
   \and INAF, Osservatorio Astrofisico di Arcetri, Largo E. Fermi 5, 50125 Firenze, Italy
   \and Dublin Institute for Advanced Studies, Astronomy \& Astrophysics Section, 31 Fitzwilliam Place, Dublin 2, Ireland
   \and Department of Astrophysics/IMAPP, Radboud University Nijmegen, PO Box 9010, NL-6500 GL Nijmegen, the Netherlands
   \and Instituto de Radioastronom\'ia y Astrof\'isica UNAM, Apartado Postal 3-72 (Xangari), 58089 Morelia, Michoac\'an, M\'exico}
   \date{Received ...; accepted ...}


  \abstract
   {}
   {We want to study the physical properties of the ionized jet emission in the vicinity of an O-type young stellar object (YSO),
   and estimate how efficient is the transfer of energy and momentum from small- to large-scale outflows.}
   {We conducted Karl G. Jansky Very Large Array (VLA) observations, at both 22 and 45\,GHz, of the compact and faint radio
   continuum emission in the high-mass star-forming region \G23, with an angular resolution between $0\farcs3$ and $0\farcs1$,
   and a thermal rms of the order of 10\,$\mu$Jy\,beam$^{-1}$. }
   {We discovered a collimated thermal (bremsstrahlung) jet emission, with a radio luminosity (L$_{\rm rad}$) of 24\,mJy\,kpc$^2$ at
   45\,GHz, in the inner 1000\,AU from an O-type YSO. The radio thermal jet has an opening angle of $44\degr$ and brings a
   momentum rate of 8$\times$10$^{-3}$\,M$_{\odot}$\,yr$^{-1}$\,km\,s$^{-1}$. By combining the new data with previous 
   observations of the molecular outflow and water maser shocks, we can trace the outflow emission from its driving source through the
   molecular clump, across more than two order of magnitude in length (500\,AU--0.2\,pc). We find that the momentum-transfer
   efficiency, between the inner jet emission and the extended outflow of entrained ambient gas, is near unity. This result suggests that
   the large-scale flow is swept-up by the mechanical force of the radio jet emission, which  originates in the inner 1000\,AU from the
   high-mass YSO.}
   {}


   \maketitle
%

\section{Introduction}

At the onset of the star formation process of early B- and O-type young stellar objects (YSOs),
faint radio continuum emission ($L_{\rm 8GHz}/L_{\rm bol}$\,$\la$10$^{-3}$), with a maximum
extent of a few 1000\,AU, is associated with radio thermal (bremsstrahlung) jets, namely, ionized
gas outflowing from the inner 100s\,AU of the newly born star (e.g., \citealt[their Fig.\,18]{Tanaka2016}).
Therefore, radio jets bring information on the outflow energetics directly produced by the star formation
process (i.e., the \emph{primary outflow}), as opposed to estimates of the molecular outflow energetics
attainable on scales greater than 0.1\,pc (through CO isotopologues typically). Large-scale molecular
outflows, or \emph{secondary outflows}, are representative of ambient gas that has been entrained by
the primary outflow. To date, only a handful of radio jets, in the vicinity 
of YSOs with $L_{\rm bol}$ greater than a few 10$^{4}$\,$L_{\odot}$,  have been studied in detail (e.g., 
\citealt{Guzman2010}, and reference therein).

Sensitive continuum observations at cm wavelengths, with an angular resolution of the order of $0\farcs1$,
are a major tool to resolve the spatial morphology of the primary outflows and measure their physical
properties (e.g., \citealt{Carrasco2010,Hofner2011,Moscadelli2016}). Here, we want to exploit the synergy
between radio continuum observations and strong H$_2$O masers, a preferred tracer of shocked gas at the base 
of (proto)stellar outflows, in order to quantify the primary outflow energetics within 1000\,AU of an O-type YSO.  
Our goal is to compare primary and secondary outflow energetics, and to estimate how efficiently the mass ejection
in the vicinity of a high-mass YSO transfers energy and momentum into large-scale motions of the clump gas.

With this in mind,  we made use of the Karl G. Jansky Very Large Array  (VLA) of the NRAO\footnote{The National Radio
Astronomy Observatory (NRAO) is a facility of the National Science Foundation operated under cooperative agreement by
Associated Universities, Inc.} to target the hot molecular core (HMC) of the star-forming region \G23, where we
previously detected compact ($\leq$\,$1\arcsec$) radio continuum emission at a level of $\la$\,1\,mJy \citep{Sanna2010}.
At a trigonometric distance of 4.6\,kpc \citep{Brunthaler2009}, \G23\, is a luminous star-forming site with a bolometric 
luminosity of 4\,$\times$\,$10^4$\,L$_{\odot}$, which harbors, at its geometrical center, an accreting
(late) O-type YSO \citep{Sanna2014}. In the past few years, we have conducted an observational campaign, with several
interferometric facilities, in order to study the environment of \G23\, from the mm to the cm wavelengths \citep{Furuya2008,
Sanna2010,Sanna2014,Sanna2015,Moscadelli2011}. Of particular interest for the present study, is the detection of a
collimated bipolar outflow, traced with SiO and CO gas emission at an angular resolution of $3\arcsec$, which lies close to the
plane of the sky at a position angle of $+58\degr$ (east of north), extends up to 0.5\,pc from the HMC
center, and brings a momentum rate of 6$\times$10$^{-3}$\,M$_{\odot}$\,yr$^{-1}$\,km\,s$^{-1}$ \citep{Sanna2014}.


\section{Observations and calibration}
\label{jvlaobs}


\begin{table}
\caption{Imaging information.\label{tabobs}}
\begin{tabular}{c c c c c c }
\hline \hline
Band & $\nu$ & HPBW & rms & S$_{\rm peak}$ & S$_{\nu}$  \\
        &  (GHz) & ($\arcsec$) & $\mu$Jy\,beam$^{-1}$ &   mJy\,beam$^{-1}$ & mJy   \\
\hline
 & & & & & \\
K & 21.7 & \multicolumn{1}{c}{0.28}  & 8   & 0.678 & 0.775 \\
Q & 44.7 & \multicolumn{1}{c}{0.15}  & 22 & 1.094 & 1.150 \\
\hline
\end{tabular}
\tablefoot{Columns\,1 and\,2 specify the observing band and the central frequency of each map, respectively. 
Column\,3 reports the restoring (circular) beam size, set equal to the geometrical average of the major and minor
axes of the dirty beam size, produced with Briggs robustness parameters 0 and 0.5 for the K and Q bands,
respectively. Columns\,4, 5, and\,6 report the rms, the peak brightness, and the flux density
integrated within the 3\,$\sigma$  contours, respectively. The maximum recoverable scale of the emission is limited
to $8\arcsec$ at K band, and $4\arcsec$ at Q band.}\\
\end{table}


We observed the star-forming region \G23\, with the B-array configuration of the VLA in the K (19.0--24.3\,GHz)
and Q  (41.6--49.8\,GHz) bands. Observations were conducted under program 15A--173 during two runs, on 2015 February
16 (Q band) and March 13 (K band). Scheduled on-source times were 1\,hr and 1.5\,hr, respectively. The phase center was set 
at R.A.\,(J2000)\,$=$\,18$^{\rm h}$34$^{\rm m}$40$\fs$29 and Dec.\,(J2000)\,$=$\,--9$\degr$00$\arcmin$38$\farcs$3.
We employed the WIDAR backend  and covered both continuum emission and a number of thermal and maser lines. In
the following, we focus on the analysis of the continuum dataset. Continuum observations were performed using the 3-bit
receivers in dual polarization mode. At K and Q bands, we employed 34 and 56 continuum spectral windows, respectively, 
each 128\,MHz wide. At both frequencies, bandpass and flux density scale calibration were obtained on 3C286; complex gain
calibration was obtained on J1832--1035. The QSO J1832--1035 had a bootstrap flux density of 0.938\,Jy and 0.576\,Jy at
22.00\,GHz and 45.00\,GHz, respectively. 

\begin{figure}
\centering
\includegraphics [angle= 0, scale= 0.68]{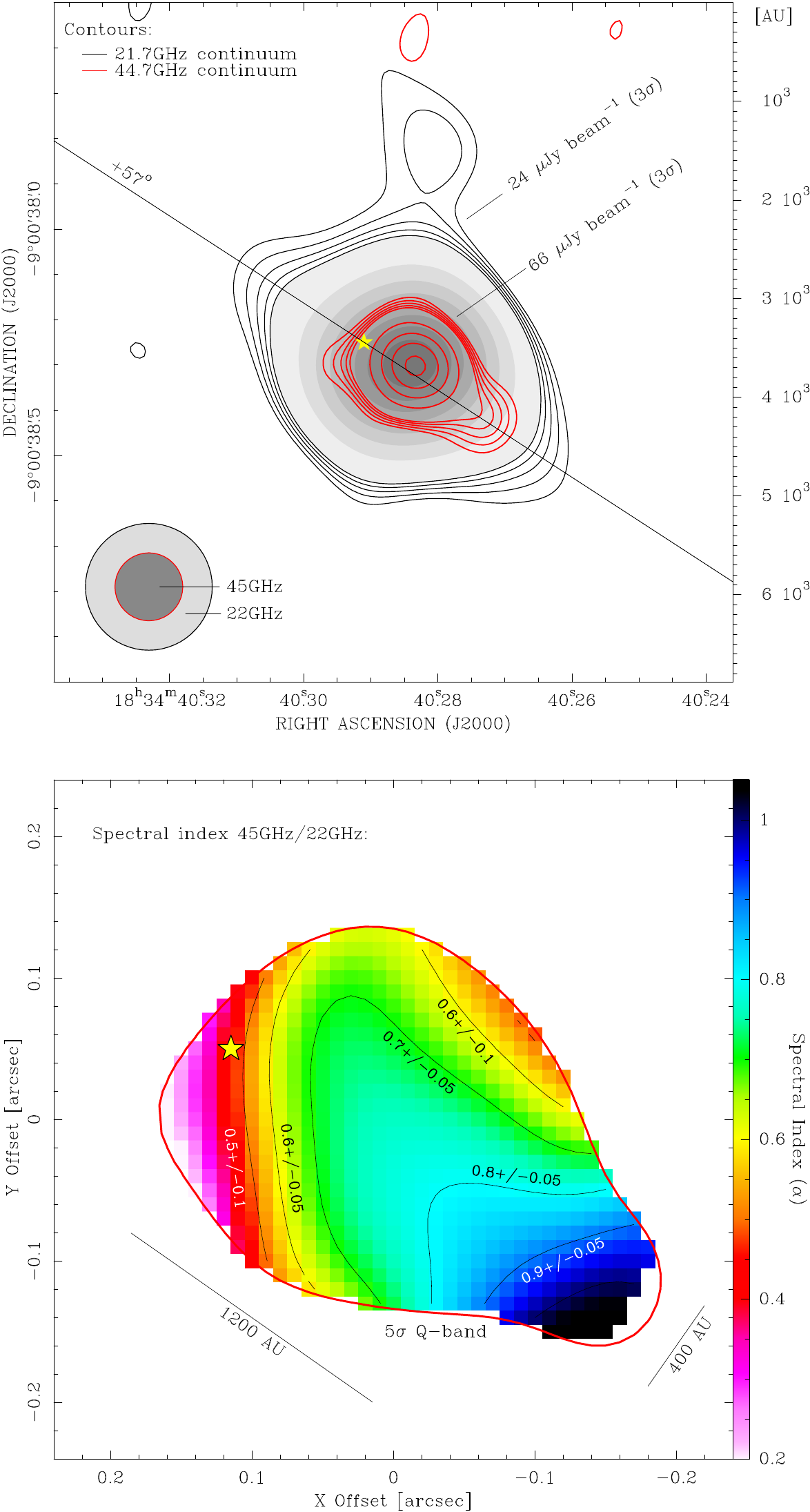}
\caption{Radio continuum emission towards \G23. 
\textbf{Upper:} overlay of the VLA maps at K (black contours \& grey scale) and Q bands
(red contours). The lower five contours start at 3\,$\sigma$ (absolute value indicated)
by 1\,$\sigma$ steps. Higher intensity levels increase by 10\,$\sigma$
steps starting from 7\,$\sigma$ (drawn in grey scale at K band). HPBWs at the 
bottom left corner. The right axis gives the linear extent of the map in AU. The star marks the HMC center, which we assume 
to be the YSO position (same for the lower panel -- see Sect.\,\ref{results}). The straight line, at
a position angle of $+57\degr$, shows the best fit to the elongation of the Q-band emission (see Sect.\,\ref{results}).
\textbf{Lower:} map of the radio spectral index (colors) between the K- and Q-band emission,
according to the righthand wedge. Spectral index levels are drawn on the map (black contours) together with their
formal uncertainty. The red thick contour shows the 5\,$\sigma$ of the Q-band map. 
Linear scales for the extension of the spectral index map are indicated.}
\label{fig1}
\end{figure}

The visibility data were calibrated with the Common Astronomy Software Applications package   (CASA v\,4.2.2), making use of the
VLA pipeline v\,1.3.1 (updated to the Perley-Butler 2013 flux density scale). After inspection and manual flagging of the dataset, we imaged
the continuum emission with the task \emph{clean} of CASA interactively. Figure\,\ref{fig1} (upper) shows an overlay of the continuum 
emission of both frequency bands. At both K and Q bands, we achieved an image rms close to the thermal noise expected from the VLA Exposure calculator
(within a factor 1.3), meaning that our maps are not dynamic-range limited. Therefore, no further self calibration was attempted. Imaging information is
summarized in Table\,\ref{tabobs}.

\begin{table*}
\caption{Radio jet properties for an isothermal conical flow with uniform ionization fraction.  \label{tabjet}}
\begin{tabular}{c c c c c c | c c c c c c c | r c}
\hline \hline
\multicolumn{6}{c}{Assumed Parameters}  & \multicolumn{7}{c}{Radio Observables} & \multicolumn{2}{c}{Jet Energetics} \\  
$\epsilon$ & q$_{\rm T}$ & q$_{\rm x}$ & $F(\alpha)$ & x$_0$ & $T$ & $\nu_{m}$ & $\nu$ & $S_{45}$\,$\times$\,$d^2$ & $\alpha$ & $\psi$ &
$i$ & $V_{\rm jet}$ & \multicolumn{1}{c}{$\dot{M}_{\rm jet}$}  & \multicolumn{1}{c}{$\dot{p}_{\rm jet}$}  \\
& & & & & (K) & (GHz) & (GHz) & (mJy\,$\times$\,kpc$^2$) & & ($\degr$)  & ($\degr$) & (km\,s$^{-1}$) & \multicolumn{1}{c}{(M$_{\odot}$\,yr$^{-1}$)} &
\multicolumn{1}{c}{(M$_{\odot}$\,yr$^{-1}$\,km\,s$^{-1}$)}  \\
\hline
 & & & & & & & & & & & & & &\\
1 & 0 & 0 & 1.5-1.1 & 1 & 10$^4$ & 50-100 & 44.7 & 1.15\,$\times$\,4.6$^2$ & 0.6-0.9  & 22$^{+8}_{-8}$ & 60-90 & 600  & 0.2-0.8$\cdot$10$^{-5}$
& 1.2-4$\cdot$10$^{-3}$ \\
& & & & & & & & & & & & 1000  & 0.4-1.4$\cdot$10$^{-5}$ & 0.4-1.4$\cdot$10$^{-2}$ \\
\textbf{1} & \textbf{0} & \textbf{0} & \textbf{1.24} & \textbf{1} & \textbf{10$^4$} & \textbf{50} & \textbf{44.7} & \textbf{1.15\,$\times$\,4.6$^2$} &
\textbf{0.75}  & \textbf{22} & \textbf{60} & \textbf{1000}  & \textbf{0.8$\cdot$10$^{-5}$} & \textbf{8$\cdot$10$^{-3}$} \\
\hline
\end{tabular}
\tablefoot{Columns 1 to 6 define the properties of the ionized gas following the formalism by \citet{Reynolds1986}. $\epsilon$,  q$_{\rm T}$, and q$_{\rm x}$ are 
the indexes of the power-law dependence, respectively, of the jet width, temperature, and ionization fraction with the jet radius ($w$\,$\propto$\,$r^{\epsilon}$,
$T$\,$\propto$\,$r^{q_{\rm T}}$, $x$\,$\propto$\,$r^{q_{\rm x}}$).  Columns 7 to 13 list the radio observables which enter into Eq.\,\ref{Mloss}. The last two
columns report the derived jet mass loss and momentum rates. The last row lists the fiducial jet parameters used in the comparison with the large-scale secondary
outflow.} 
\end{table*}


\section{Results}\label{results}

The radio continuum emission extends within 2000--3000\,AU of the HMC center (star symbol in Fig.\,\ref{fig1}).
The HMC center is defined by the peak position of a high-excitation (E$_{\rm low}$ of 363\,K) CH$_3$OH thermal line at the systemic
velocity of the core \citep[their Table\,2]{Sanna2014}.  In the upper panel of Fig.\,\ref{fig1}, the Q-band emission shows a spatial
morphology which is stretched along the NE--SW direction, with the bulk of the emission to the SW of the HMC center.  On the other hand,
the K-band emission shows a boxy morphology with the NE--SW diagonal line coincident with the Q-band major axis of the emission. The peak
positions of the K- and Q-band maps  spatially overlap at R.A.\,(J2000)\,$=$\,18$^{\rm h}$34$^{\rm m}$40$\fs$284  and
Dec.\,(J2000)\,$=$\,--9$\degr$00$\arcmin$38$\farcs$30, with an uncertainty of 30\,mas in each coordinate.
Despite having imaged a field-of-view of 0.2\,pc around the radio continuum peak, we did not detect any other radio continuum source
exceeding a brightness of 40\,$\mu$Jy\,beam$^{-1}$ at K band (5\,$\sigma$).

In Fig.\,\ref{fig1} (lower), we present a spectral index ($\alpha$) map of the continuum emission between 22 and 45\,GHz,
obtained with the task \emph{clean} of CASA by multi-frequency synthesis cleaning, and assuming a constant spectral slope
\citep{Rau2011}. The spectral index is computed only where the average intensity map produced by \emph{clean} has SNR$\geq$10.
For comparison, we also draw the 5\,$\sigma$ contour of the Q-band emission. In this map, the levels of spectral index (and the
corresponding 1\,$\sigma$ uncertainties) are also shown, in steps of 0.1.  If we leave out the outer edges of the spectral index map,
where the uncertainties increase to $>$\,0.1, the spectral index changes smoothly between 0.6--0.9 along the NE-SW elongation of
the continuum emission. This range of values is consistent with the radio continuum emission being produced by thermal bremsstrahlung
from  a stellar wind  \citep{Panagia1975,Reynolds1986,Anglada1998}. It is worth noting that the contribution of the dust emission
at 7\,mm should be negligible, since the 7\,mm flux, extrapolated from our previous measurements at 3 and 1\,mm
\citep{Furuya2008,Sanna2014}, is less than 10\,\%  of the total Q-band flux.   

In order to better discriminate the spatial morphology of the Q-band emission, we removed from the visibility dataset a point-like source model
of 0.9\,Jy, set at the peak position of the Q-band map. We then cleaned the new visibility dataset with the same parameters used to produce the
Q-band map of  Fig.\,\ref{fig1}. The result of this procedure is shown in the right panel of Fig.\,\ref{fig2} (hereafter, the residual map). The NE--SW
extended component of the Q-band emission approaches the star position with no significant changes in shape. By fitting a two-dimensional
Gaussian to the residual map, the position angle of its major axis is $+57\degr\pm3\degr$. This best fit is drawn in Figs.\,\ref{fig1} and\,\ref{fig2}.
The orientation of the Q-band emission closely matches the direction ($+58\degr$) of the CO\,(2--1) outflow estimated at a scale $\ga$\,0.1\,pc
 (e.g., Fig.\,\ref{fig2}, left). Combining this evidence with the spectral index analysis, we infer that the radio continuum at Q band traces a
 collimated thermal jet, with its peak emission pinpointing a strong jet knot.

\section{Discussion}\label{discussion}

\begin{figure*}
\centering
\includegraphics [angle= 0, scale= 0.9]{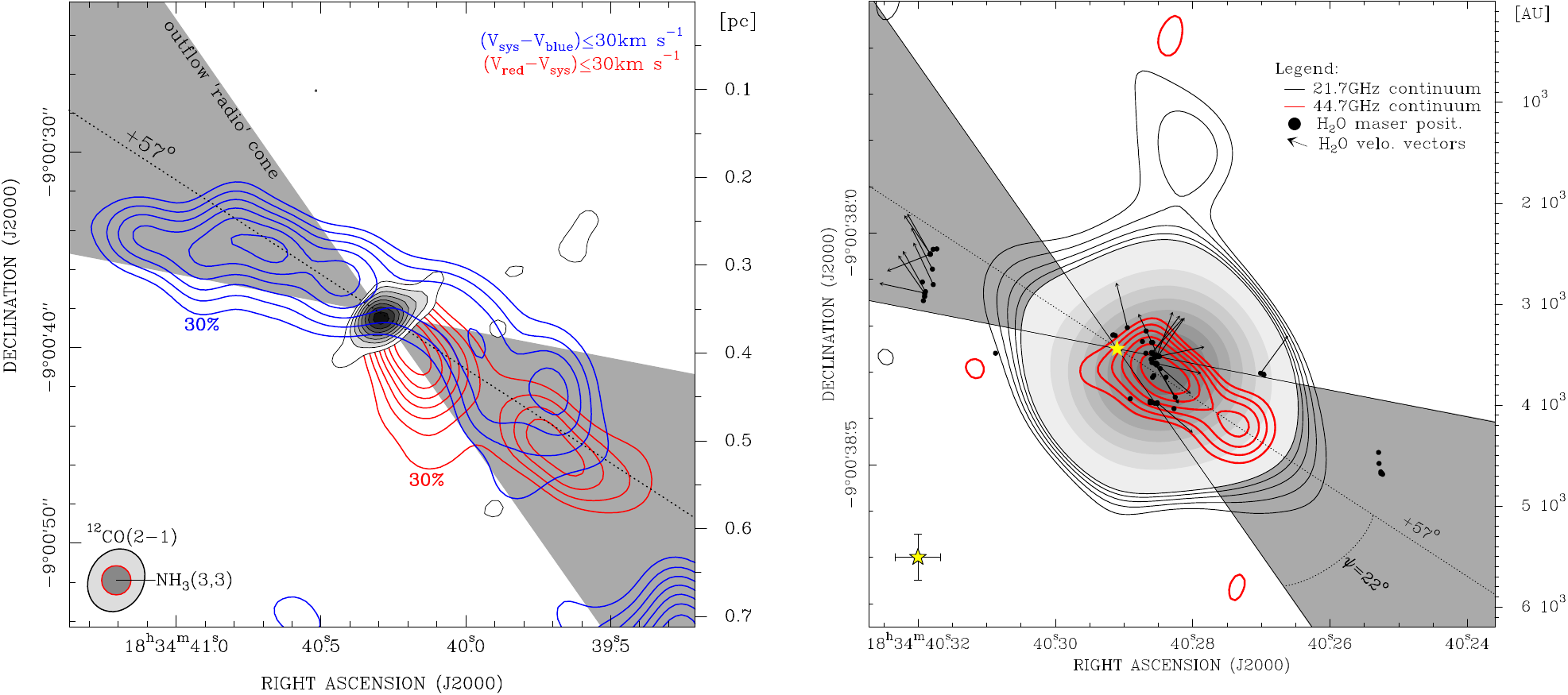}
\caption{Comparison of the outflow tracers across different scales (linear extent of each map on the right axes). 
\textbf{Right:} overlay of the K-band emission (same as in Fig.\,\ref{fig1}) with the Q-band residual map (see Sect.\,\ref{results}).
Residual map contours start at 3\,$\sigma$ by 1\,$\sigma$ steps. Restoring beams and star symbol as in Fig.\,\ref{fig1}. The uncertainty 
of the star position is shown in the bottom left corner. For comparison, we overlay the distribution (black spots) and the velocity vectors
(black arrows) of the 22.2 GHz H$_2$O maser cloudlets \citep{Sanna2010}. The grey cone marks the boundary of the opening angle
($2\,\psi$) of the radio jet emission (see Sect.\,\ref{discussion}).
\textbf{Left:} Comparison between the spatial distribution of the  outflow entrained $^{12}$CO gas \citep{Sanna2014} with
the opening angle of the radio jet emission (grey shadow). Blue and red contours (10\,\% steps starting from 30\,\% of the peak emission) 
show the distribution of the blue- and red-shifted CO gas emission, with LSR velocities up to 30\,\kms\ from the systemic velocity. Grey
contours at the center of the bipolar outflow show the distribution of warm NH$_3$ gas emission \citep{Codella1997}.  For the NH$_3$ map,
contours start at 4\,$\sigma$ by 1\,$\sigma$ steps. Synthesized HPBWs are drawn at the bottom left corner.}
\label{fig2}
\end{figure*}

In Sect.\,\ref{results}, we have addressed the nature of the radio continuum emission and shown that the Q-band flux is dominated by 
thermal jet emission. In the following, we study the physical properties of this radio jet in detail. We assume the HMC center as the best 
guess for the YSO position. This position also coincides with the origin of the velocity fields measured with the CH$_3$OH and H$_2$O 
masers cloudlets \citep{Sanna2010,Sanna2015}, within an uncertainty of $\pm50$\,mas. This uncertainty does not affect the following
calculations significantly.  

First, we make use of the Q-band residual map (Fig.\,\ref{fig2}, right) to quantify the degree of collimation of the radio jet. At the positions
of the two local peaks of the residual map, we estimate the FWHM of the jet emission  ($0\farcs145$ and $0\farcs160$
for the stronger and fainter peak, respectively) perpendicular to the jet direction (i.e., the Q-band major axis). The jet semi-opening angle ($\psi$), of
 $22\degr$\,$\pm$\,$8\degr$, is defined by the average (semi-)angle evaluated from the star position to each FWHM. In the right panel of Fig.\,\ref{fig2},
we draw the boundary of the radio jet opening angle with a grey cone. In the left panel of Fig.\,\ref{fig2}, we compare the same ``radio cone'' 
(grey shadow) with the spatial distribution of the blue- and red-shifted ambient gas, and show that it comprises about 70\% of the CO\,(2--1)
outflow emission.

In order to quantify the energetics of the radio jet, we derive the mass loss rate of the ionized gas following the analytic treatment by
\citet{Reynolds1986}. Hereafter, we make use of the formalism introduced by these authors for a direct comparison. This approach
describes the behavior of a collimated conical flow (or jet) without assuming any specific ionization mechanism.  We rewrite Eq.\,19 of
\citet{Reynolds1986} in a convenient form\footnote{With respect to Eq.\,19 of \citet{Reynolds1986}, we have assumed a pure hydrogen 
jet, $\mu/m_{\rm p}$\,$=$\,1, where $\mu$ is the mean particle mass per hydrogen atom, and $m_{\rm p}$ is the proton mass. Also,
we have replaced the parameter $\theta_0$, defined by the ratio of the FWHM of the jet (2$\omega_0$) and its distance from the star ($r_0$),
with the jet semi-opening angle, $\psi$ ($\theta_0$\,$=$\,2\,$\tan\,\psi$).}:  
{\small
\begin{displaymath}
\left(\frac{\dot{M}}{\rm 10^{-5}\,M_{\odot}\,yr^{-1}}\right)=
1.6\cdot10^{-2} \times F^{-3/4} \times x_0^{-1} \times \left(\frac{T}{\rm 10^4\,K}\right)^{-3/40}
\end{displaymath}
\begin{displaymath}
\times \left(\frac{\nu}{\rm 10\,GHz}\right)^{-(3/4)\alpha} \times \left(\frac{\nu_{\rm m}}{\rm 10\,GHz}\right)^{+(3/4)\alpha-9/20}
\times \left(\frac{S_{\nu} \times d^2}{\rm mJy \times kpc^2}\right)^{3/4}
\end{displaymath}
\begin{equation}\label{Mloss}
\times \left(\tan \,\psi\right)^{3/4} \times  \left({\sin}\,i\right)^{-1/4} \times \left(\frac{V_{\rm jet}}{\rm 100\,km\,s^{-1}}\right).
\end{equation}
}
The first three terms define the conditions of the ionized gas. The last six terms are the observables of the radio jet emission. 
$F$ is an index for the jet optical depth, $x_0$ is the hydrogen ionization fraction, and T is the ionized gas temperature.  
The radio observables are the flux ($S_{\rm \nu}$), frequency ($\nu$), and spectral index of the emission ($\alpha$), the turn-over
frequency of the radio jet spectrum ($\nu_{\rm m}$), the semi-opening angle of the radio jet ($\psi$), its inclination with the line
of sight ($i$), and the expanding velocity of the ionized gas ($V_{\rm jet}$). In Table\,\ref{tabjet}, we list the values (or ranges) used to
solve Eq.\,\ref{Mloss}. For a 20\,M$_{\odot}$ star, \citet[their Fig.\,3]{Tanaka2016} show that the outflowing gas, in the inner
1000\,AU, has a uniform ionization fraction near unity, and an isothermal  gas temperature of $10^4$\,K. For a pure conical flow,
Reynolds' model is completely determined by these two conditions and the spectral index value of the radio jet emission. $F$ can be
directly computed with Eqs.\,15 and\,17 of \citet{Reynolds1986}. 

In Table\,\ref{tabjet}, we make use of the Q-band observations to fix $\nu$, $S_{\rm \nu}$, $\alpha$, and $\psi$.
The minimum inclination ($i$) of the outflow axis is set to 60$\degr$ based on our  SMA observations at 1\,mm \citep{Sanna2014}.
A  lower limit to the turn-over frequency ($\nu_{\rm m}$) is provided by the higher frequency of the Q-band observations (50\,GHz).
It is worth noting that, the value of mass loss rate from  Eq.\,\ref{Mloss}  changes by less than $10\%$ for any variation of $i$
and $\nu_{\rm m}$ in the ranges 60$\degr$--90$\degr$ and 50--100\,GHz, respectively. 

To date, radio jet velocities, between 200--1400\,\kms,  have been directly measured towards a couple of targets only, as this requires a
long-term monitoring of the jet morphology \citep{Marti1995,Curiel2006,Guzman2016,Kamenetzky2016}. Here, we show
that we can provide an estimate of the radio jet motion ($V_{\rm jet}$) from the velocity vectors of the H$_2$O maser cloudlets  ($V_{\rm H_2O}$).
The 22.2\,GHz H$_2$O maser transition is a tracer of shocked ambient gas at typical densities of $\sim$10$^8 \rm cm^{-3}$ (e.g.,
\citealt{Hollenbach2013}, their Fig.\,15). Figure\,\ref{fig2} (right) shows that the H$_2$O maser cloudlets mainly lie inside the jet cone.
Close to the HMC center, the H$_2$O maser velocity field likely underlines a bow-shock, which would account for the
velocity vectors either parallel or perpendicular to the jet axis. The shocked layer of maser emission, further to the NE, shows velocity
vectors well aligned with the outflow direction. Balancing the ram pressure between the ionized jet and the swept-up H$_2$O masing gas,
at a shock interface, we can derive the incoming jet velocity as a function of the H$_2$O maser velocity along the jet direction, the
pre-shocked density of the H$_2$O gas ($n_{\rm H_2O}$), and the ionized gas density ($n_{\rm jet}$):
{\small
\begin{equation}\label{vjet}
\left(\frac{V_{\rm jet}}{\rm 100\,km\,s^{-1}}\right) \approx \left(\frac{V_{\rm H_2O}}{\rm 10\,km\,s^{-1}}\right)
\times \left(\frac{n_{\rm H_2O}}{\rm 10^8 cm^{-3}}\right)^{1/2} \times \left(\frac{n_{\rm jet}}{\rm 10^6 cm^{-3}}\right)^{-1/2}
\end{equation}
}
Eq.\,\ref{vjet} is accurate to better than $10\%$ provided that the density ratio between the pre-shocked and ionized gas exceeds a factor of $100$.
\citet{Tanaka2016} show that, in the inner 1000\,AU, an ionized gas density of $10^6 \rm cm^{-3}$ is most appropriate.
Given a maximum H$_2$O maser velocity of 60\,\kms\, \citep{Sanna2010}, Eq.\,\ref{vjet} implies a jet velocity of 600\,\kms\,  at a few 100\,AU
from the YSO. According to \citet{Reynolds1986}, if the flow is isothermal and uniformly ionized, a spectral index value higher than 0.6 corresponds 
to a radio jet with velocities slowly increasing with distance from the star. Taking into account the power-law dependence of the jet velocity with the 
jet radius, we estimate a velocity in excess of 1000\,\kms\, at the southwestern tip of the Q-band map, at a distance of 1700\,AU from the YSO.

Finally, we draw a comparison between the mechanical properties of the radio jet and the molecular outflow \citep[Table\,5]{Sanna2014}.
The maximum and minimum values of mass loss ($\dot{M}_{\rm jet}$) and momentum rate ($\dot{p}_{\rm jet}$) of the radio jet, for the possible
combinations of radio observables,  are listed in the last columns of Table\,\ref{tabjet}, for a minimum velocity of 600\,\kms, and an outer jet velocity
of 1000\,\kms.  The last row of Table\,\ref{tabjet} lists the fiducial jet parameters used in the following discussion. Since we only detect the SW lobe 
of the jet emission at Q-band, we multiply the mechanical properties of the radio jet by a factor of 2, to compare with the large-scale bipolar outflow.
Asymmetric radio continuum  emission is frequently observed around YSOs of different masses (e.g., \citealt{Hofner2007,Johnston2013}),  and may
be due to density inhomogeneity of the protostellar environment. Figure\,\ref{fig2} allows one to trace backward the outflow emission, down to its
driving source, over more than two order of magnitude of distance from the HMC center (0.2\,pc--500\,AU). We can interpret this result as the
evidence for a single object dominating the clump dynamics. If the large-scale ambient gas is accelerated by the jet, the total (time-averaged)
momentum released by the radio jet into the clump gas, $p_{\rm jet}$\,$=$\,160\,M$_{\odot}$\,km\,s$^{-1}$, can be determined by multiplying
the jet momentum rate by the dynamical time of the molecular outflow, $t_{\rm dyn}$\,$=$\,2\,$10^4$\,yr. The comparison of the momentum
provided by the radio jet with that of the molecular outflow, 116\,M$_{\odot}$\,km\,s$^{-1}$, implies a momentum-transfer efficiency near unity.
This result provides evidence that the large-scale flow is swept-up by the radio thermal jet, which originates in the inner 1000\,AU from the high-mass YSO.

\begin{acknowledgements}

Comments from an anonymous referee, which helped improving our paper, are gratefully acknowledged.
Financial support by the Deutsche Forschungsgemeinschaft (DFG) Priority Program 1573 is gratefully acknowledged.
The authors thank S.\,Dzib, P.\,Hofner, R.\,Kuiper, and A.\,K\"{o}lligan for fruitful discussions in preparation. 

\end{acknowledgements}


\bibliographystyle{aa}
\bibliography{asanna1708}

\begin{thebibliography}{23}
\expandafter\ifx\csname natexlab\endcsname\relax\def\natexlab#1{#1}\fi

\bibitem[{{Anglada} {et~al.}(1998){Anglada}, {Villuendas}, {Estalella},
  {Beltr{\'a}n}, {Rodr{\'{\i}}guez}, {Torrelles}, \& {Curiel}}]{Anglada1998}
{Anglada}, G., {Villuendas}, E., {Estalella}, R., {et~al.} 1998, \aj, 116, 2953

\bibitem[{{Brunthaler} {et~al.}(2009){Brunthaler}, {Reid}, {Menten}, {Zheng},
  {Moscadelli}, \& {Xu}}]{Brunthaler2009}
{Brunthaler}, A., {Reid}, M.~J., {Menten}, K.~M., {et~al.} 2009, \apj, 693, 424

\bibitem[{{Carrasco-Gonz{\'a}lez} {et~al.}(2010){Carrasco-Gonz{\'a}lez},
  {Rodr{\'{\i}}guez}, {Anglada}, {Mart{\'{\i}}}, {Torrelles}, \&
  {Osorio}}]{Carrasco2010}
{Carrasco-Gonz{\'a}lez}, C., {Rodr{\'{\i}}guez}, L.~F., {Anglada}, G., {et~al.}
  2010, Science, 330, 1209

\bibitem[{{Codella} {et~al.}(1997){Codella}, {Testi}, \&
  {Cesaroni}}]{Codella1997}
{Codella}, C., {Testi}, L., \& {Cesaroni}, R. 1997, \aap, 325, 282

\bibitem[{{Curiel} {et~al.}(2006){Curiel}, {Ho}, {Patel}, {Torrelles},
  {Rodr{\'{\i}}guez}, {Trinidad}, {Cant{\'o}}, {Hern{\'a}ndez}, {G{\'o}mez},
  {Garay}, \& {Anglada}}]{Curiel2006}
{Curiel}, S., {Ho}, P.~T.~P., {Patel}, N.~A., {et~al.} 2006, \apj, 638, 878

\bibitem[{{Furuya} {et~al.}(2008){Furuya}, {Cesaroni}, {Takahashi}, {Codella},
  {Momose}, \& {Beltr{\'a}n}}]{Furuya2008}
{Furuya}, R.~S., {Cesaroni}, R., {Takahashi}, S., {et~al.} 2008, \apj, 673, 363

\bibitem[{{Guzm{\'a}n} {et~al.}(2010){Guzm{\'a}n}, {Garay}, \&
  {Brooks}}]{Guzman2010}
{Guzm{\'a}n}, A.~E., {Garay}, G., \& {Brooks}, K.~J. 2010, \apj, 725, 734

\bibitem[{{Guzm{\'a}n} {et~al.}(2016){Guzm{\'a}n}, {Garay}, {Rodr{\'{\i}}guez},
  {Contreras}, {Dougados}, \& {Cabrit}}]{Guzman2016}
{Guzm{\'a}n}, A.~E., {Garay}, G., {Rodr{\'{\i}}guez}, L.~F., {et~al.} 2016,
  \apj, 826, 208

\bibitem[{{Hofner} {et~al.}(2007){Hofner}, {Cesaroni}, {Olmi},
  {Rodr{\'{\i}}guez}, {Mart{\'{\i}}}, \& {Araya}}]{Hofner2007}
{Hofner}, P., {Cesaroni}, R., {Olmi}, L., {et~al.} 2007, \aap, 465, 197

\bibitem[{{Hofner} {et~al.}(2011){Hofner}, {Kurtz}, {Ellingsen}, {Menten},
  {Wyrowski}, {Araya}, {Loinard}, {Rodr{\'{\i}}guez}, \&
  {Cesaroni}}]{Hofner2011}
{Hofner}, P., {Kurtz}, S., {Ellingsen}, S.~P., {et~al.} 2011, \apjl, 739, L17

\bibitem[{{Hollenbach} {et~al.}(2013){Hollenbach}, {Elitzur}, \&
  {McKee}}]{Hollenbach2013}
{Hollenbach}, D., {Elitzur}, M., \& {McKee}, C.~F. 2013, \apj, 773, 70

\bibitem[{{Johnston} {et~al.}(2013){Johnston}, {Shepherd}, {Robitaille}, \&
  {Wood}}]{Johnston2013}
{Johnston}, K.~G., {Shepherd}, D.~S., {Robitaille}, T.~P., \& {Wood}, K. 2013,
  \aap, 551, A43

\bibitem[{{Marti} {et~al.}(1995){Marti}, {Rodriguez}, \&
  {Reipurth}}]{Marti1995}
{Marti}, J., {Rodriguez}, L.~F., \& {Reipurth}, B. 1995, \apj, 449, 184

\bibitem[{{Moscadelli} {et~al.}(2016){Moscadelli}, {S{\'a}nchez-Monge},
  {Goddi}, {Li}, {Sanna}, {Cesaroni}, {Pestalozzi}, {Molinari}, \&
  {Reid}}]{Moscadelli2016}
{Moscadelli}, L., {S{\'a}nchez-Monge}, {\'A}., {Goddi}, C., {et~al.} 2016,
  \aap, 585, A71

\bibitem[{{Moscadelli} {et~al.}(2011){Moscadelli}, {Sanna}, \&
  {Goddi}}]{Moscadelli2011}
{Moscadelli}, L., {Sanna}, A., \& {Goddi}, C. 2011, \aap, 536, A38

\bibitem[{{Panagia} \& {Felli}(1975)}]{Panagia1975}
{Panagia}, N. \& {Felli}, M. 1975, \aap, 39, 1

\bibitem[{{Rau} \& {Cornwell}(2011)}]{Rau2011}
{Rau}, U. \& {Cornwell}, T.~J. 2011, \aap, 532, A71

\bibitem[{{Reynolds}(1986)}]{Reynolds1986}
{Reynolds}, S.~P. 1986, \apj, 304, 713

\bibitem[{{Rodr{\'{\i}}guez-Kamenetzky}
  {et~al.}(2016){Rodr{\'{\i}}guez-Kamenetzky}, {Carrasco-Gonz{\'a}lez},
  {Araudo}, {Torrelles}, {Anglada}, {Mart{\'{\i}}}, {Rodr{\'{\i}}guez}, \&
  {Valotto}}]{Kamenetzky2016}
{Rodr{\'{\i}}guez-Kamenetzky}, A., {Carrasco-Gonz{\'a}lez}, C., {Araudo}, A.,
  {et~al.} 2016, \apj, 818, 27

\bibitem[{{Sanna} {et~al.}(2014){Sanna}, {Cesaroni}, {Moscadelli}, {Zhang},
  {Menten}, {Molinari}, {Caratti o Garatti}, \& {De Buizer}}]{Sanna2014}
{Sanna}, A., {Cesaroni}, R., {Moscadelli}, L., {et~al.} 2014, \aap, 565, A34

\bibitem[{{Sanna} {et~al.}(2010){Sanna}, {Moscadelli}, {Cesaroni}, {Tarchi},
  {Furuya}, \& {Goddi}}]{Sanna2010}
{Sanna}, A., {Moscadelli}, L., {Cesaroni}, R., {et~al.} 2010, \aap, 517, A78

\bibitem[{{Sanna} {et~al.}(2015){Sanna}, {Surcis}, {Moscadelli}, {Cesaroni},
  {Goddi}, {Vlemmings}, \& {Caratti o Garatti}}]{Sanna2015}
{Sanna}, A., {Surcis}, G., {Moscadelli}, L., {et~al.} 2015, \aap, 583, L3

\bibitem[{{Tanaka} {et~al.}(2016){Tanaka}, {Tan}, \& {Zhang}}]{Tanaka2016}
{Tanaka}, K.~E.~I., {Tan}, J.~C., \& {Zhang}, Y. 2016, \apj, 818, 52

\end{thebibliography}






\end{document}